# Title: The Use of Generative Search Engines for Knowledge Work and Complex Tasks

**Authors:** Siddharth Suri[1]*†, Scott Counts[1]*†, Leijie Wang[3]‡, Chacha Chen[4]‡, Mengting Wan[2], Tara Safavi[1], Jennifer Neville[1], Chirag Shah[3]§, Ryen W. White[1], Reid Andersen[2], Georg Buscher[2], Sathish Manivannan[2], Nagu Rangan[2], Longqi Yang[2]

**Affiliations:**

[1] Microsoft Research; Redmond, WA USA.

[2]Microsoft Corporation; Redmond, WA USA.

[3]University of Washington; Seattle, WA USA.

[4]University of Chicago; Chicago, IL USA.

*Corresponding authors. Email: {suri, counts}@microsoft.com

†These authors contributed equally to this work.

‡These authors were interns at Microsoft Research when this work was done.

§This author was a visiting researcher at Microsoft Research when this work was done.

**Abstract:** Until recently, search engines were the predominant method for people to access online information. The recent emergence of large language models (LLMs) has given machines new capabilities such as the ability to generate new digital artifacts like text, images, code etc., resulting in a new tool, a generative search engine, which combines the capabilities of LLMs with a traditional search engine. Through the empirical analysis of Bing Copilot (Bing Chat), one of the first publicly available generative search engines, we analyze the types and complexity of tasks that people use Bing Copilot for compared to Bing Search. Findings indicate that people use the generative search engine for more knowledge work tasks that are higher in cognitive complexity than were commonly done with a traditional search engine.

**One-Sentence Summary:** People use generative search engines primarily for knowledge work tasks that are more complex than tasks done via a traditional search engine.

**Main Text:** Search engines are one of, if not, the predominant methods for accessing online information. The recent rise of large language models (LLMs) has given us a new kind of search engine, a generative search engine, that people can use not just to access information but to generate new digital artifacts such as information, text, and code as well as summaries, comparisons, and analyses of those artifacts. In fact, LLMs have made machines capable of doing a variety of tasks such as coding, mathematical reasoning, and image generation *(1)*, that were previously difficult, if not impossible, for a machine to do on its own. Furthermore, the potential gains from this new technology are widespread as people around the world have access through openly available LLM powered AI systems such as ChatGPT, Bing Copilot (formerly Bing Chat), and Gemini (formerly Bard) among others. With any new technology, especially one as widespread and popular as LLMs, it is important to analyze how people are using it and what they are using it for.

Thus, the overarching question this work seeks to address is how do people use generative search engines and how does that usage compare to traditional search engines? We will examine this question in the context of a generative search engine, Bing Copilot *(2)*, which marries the new capabilities of LLMs with a traditional search engine which can retrieve information from the web. We will operationalize this broad question in two ways. First, to assess "the what" we will analyze the topical domains of Bing Copilot conversations compared to Bing Search *(3)* sessions. Second, regarding "the how" we will compare the complexity of the tasks people try to use Bing Copilot for with the tasks people use Bing Search for.

Early research on LLM use indicates that humans can take advantage of their capabilities to improve task performance, especially among non-experts, in programming *(4)*, writing *(5)*, consulting *(6)*, customer support *(7)*, and general creativity *(8)*. One thing these studies have in common is that each one is an experiment where an experimenter gave a participant a knowledge-work task in a professional context. Knowledge work is an important part of the economy, with representative sectors such the Professional, Scientific, and Technical Services sector employing approximately 10.5 million workers and accounting for over 2.6 trillion dollars of output in the United States alone in 2022 *(9)*. In an arguably surprising turn, knowledge work has become the consensus frontrunner form of labor to see significant impact from AI *(10)* which even further motivates understanding how and for what people use LLMs in the wild. In this work we will examine how people use Bing Copilot, in the wild, for their own tasks. We will show that, in fact, knowledge work tasks which are more complex than tasks typically seen in traditional search engine usage are a primary use case for users of Bing Copilot.

## Results

We created a dataset of 80000 randomly sampled, de-identified Bing Copilot conversations by selecting 10000 conversations at random for each of eight consecutive weeks (weeks starting May 28, 2023 to ending July 22, 2023), also ensuring that each conversation came from a unique user. Of the 80000 conversations we sampled, 112 conversations we unusable, mostly due to incoherent text, leaving 79888 conversations. Using a methodology described in the SM and *(11)* we used GPT-4 to create labels and domain definitions and classify each Bing Copilot conversation into one of 25 topical domains (plus 'Other'). While we leave the details of this algorithm to the SM, we note that it is an unsupervised algorithm that generates the taxonomy of these data organically in a bottom-up fashion while avoiding biasing the output with our own assumptions as to what may or may not be in the data. Approximately 3.3% of conversations were labeled with nonconforming labels by our prompt-based classifiers by either this domain classifier or during one of the later classifiers, leaving a final dataset of 77303 fully classified conversations shown in Figure 1 (top, left). Recently there have been a number of papers demonstrating LLMs ability in a variety of natural language processing tasks including classification (12,13,14), nonetheless we provide our own human validation for this specific data set and task in the SM.

As a baseline comparison we ran the same domain taxonomy generation process on a comparable data set of Bing Search sessions sampled the same way over the same period (approximately 10000 randomly sampled search sessions from logged in users for each of eight weeks starting May 28, 2023). Nonconforming domain or other classification labels affected about 1.6% of sessions, leaving a final dataset of 78701 Bing Search sessions. When the consumer version of Bing Copilot was first launched, which is the version we study, it required users to be logged into their Microsoft accounts; thus, to make the data sets comparable, we required users to be logged into their Microsoft accounts for both the Bing Copilot and Bing Search data sets. In the top 10 most common Bing Search domains are social media, shopping, communication, entertainment and gaming (see Figure 1 top). Of those only gaming and entertainment exist in the top 10 of Bing Copilot domains. By contrast, in the top 10 most common domains of Bing Copilot are business and economics, translation and language learning, and biology and medicine (see Figure 1 bottom) which either do not exist in the Bing Search domains or exist outside of the top 10. Note, a common category in the Bing Search taxonomy is "Other", since many search queries are just one word and lacking context, it is often difficult to infer the domain of the query thus such queries get labelled "Other".

To facilitate a more direct comparison between Bing Copilot and Bing Search usage we additionally categorized all the Bing Search sessions using the Bing Copilot taxonomy in Figure 1 bottom. Among the top 16 Bing Copilot categories, making up 86.4% percent of total conversation sample, five categories have roughly comparable representation between Bing Copilot and Chat (Business and economics; Web development; History and culture; Education and learning; Food and drink), nine categories have between roughly 1.5 times and double the percentage in Bing Copilot compared to Bing Search (Translation and language learning; Creative Writing and editing; Biology and medicine; Programming and scripting; Academic Writing and editing; Small talk and chatbot; Engineering and design; Law and politics; Data analysis and visualization), and three categories have between roughly 1.5 times and double the percentage in Bing Search compared to Bing Copilot (Gaming and entertainment; Travel and tourism; Fashion and beauty).

More specifically, many of the top domains for Bing Copilot, and the domains where there are more Bing Copilot conversations than Bing Search sessions, appear related to knowledge work. We define knowledge work *(15)* as work that concerns the creation, handling, and distribution of information and knowledge products, and that involved non-routine tasks and uses creative and analytical thinking in convergent and divergent ways *(16, 17)*. To highlight this trend, we categorize each domain as knowledge work if it aligns with these four U.S. Bureau of Labor Statistics (BLS) employment sectors: Professional, Scientific, and Technical Services; Information; Finance and Insurance; and Educational Services *(18)* and color code Figure 1 accordingly. Overall, 72.9% of Bing Copilot conversations are in knowledge work domains whereas only 37% of Bing Search sessions are in knowledge work domains.

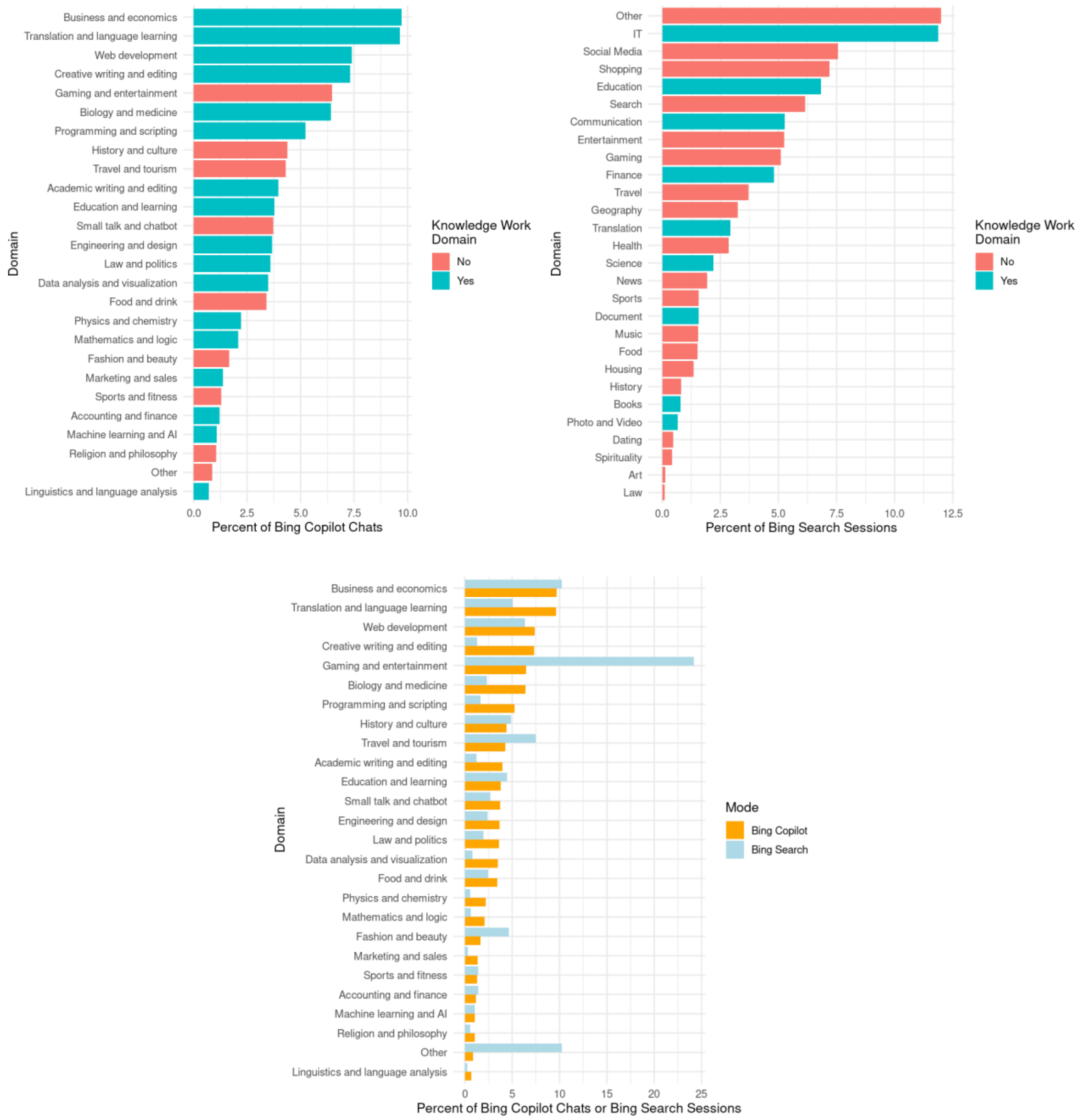

**Figure 1.** Percent of topical domains from a random sample of approximately 80000 Bing Copilot chat conversations (top, left) and 80000 Bing Search sessions (top, right). To facilitate comparison between Bing Search and Bing Copilot, bottom shows the fraction of Bing Search sessions categorized using the Bing Copilot domains.

## Task Complexity

Given that generative search engines enable users to understand, analyze, and create digital artifacts, do we see evidence of these behaviors in Bing Copilot? We analyzed the cognitive complexity of the main task that users were trying to accomplish using Bing Copilot and Bing Search at the conversation and search session levels. More specifically, we asked what would be the cognitive complexity of those tasks be if a human were to complete them without the use of an AI?

We used GPT-4 to classify the main task in each conversation and session according to Anderson and Krathwohl's Taxonomy of learning domains which defines six categories from lowest complexity to highest: Remember, Understand, Apply, Analyze, Evaluate, and Create. We chose this taxonomy because it was designed to categorize the learning tasks for *humans* and to find the commonalities between them regardless of topical domain (e.g. math and art) which is important in our context as well. Furthermore, it has previously been used in the context of information retrieval and search engine usage showing that users of traditional search engines exhibit higher levels of activity such as more searching, clicking, and bookmarking, and also require longer completion times when performing more complex tasks *(19, 20, 21, 22, 23)*.

We faithfully implemented Anderson and Krathwohl's Taxonomy into a classification prompt (see SM for the prompt and validation). Figure 2 shows the distribution of Bing Copilot conversations (n=79471) and Bing Search sessions (n=79104) across cognitive complexity categories. Only 13.4% of Bing Search sessions fall into the higher complexity categories of Apply, Analyze, Evaluate, or Create, whereas 37.0% of Bing Copilot conversations belong to those four complexity categories which marks a large qualitative shift in the use cases of the two systems. Thus, not only are people using Bing Copilot for more knowledge work, but they are also using them for more complex tasks.

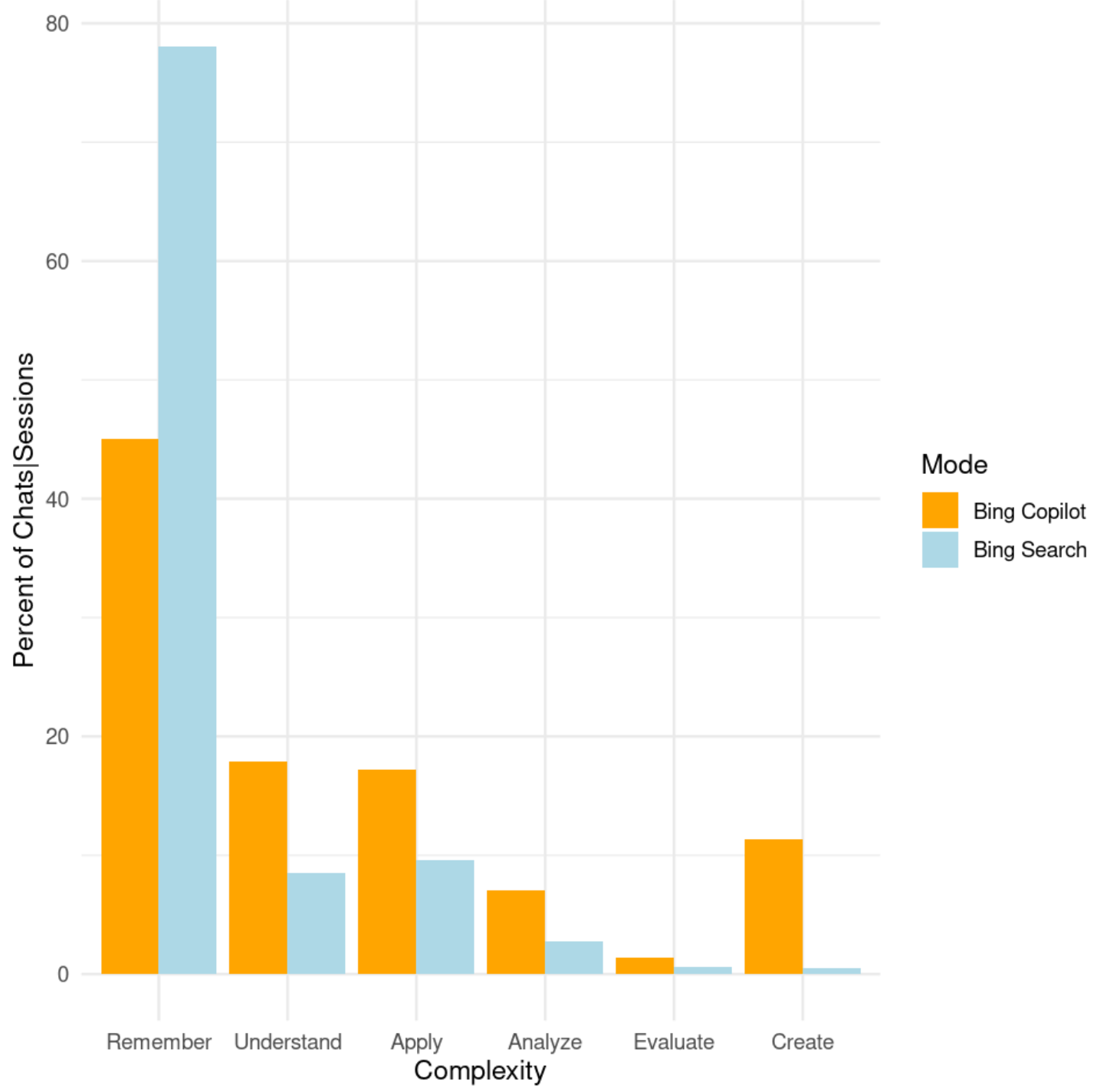

**Figure 2.** Percentage of tasks within each complexity category for Bing Copilot conversations and Bing Search Sessions.

Given that people use Bing Copilot for more complex tasks, we next test if they are more satisfied when a more complex task is completed. To do this we created a prompt that rates the extent to which a user was satisfied with the conversation with Bing Copilot. Within a Bing Copilot conversation users can vote on a conversation with a thumbs up or thumbs down and we validated our satisfaction prompt using these votes (see SM for details and (*24)* for validation). Scores on this satisfaction scale can range from -100 (maximally unsatisfied) to 100 (maximally satisfied), but in practice satisfaction for most Bing Copilot conversations falls between -10 and 30. We also created a prompt which analyzes a conversation and determines if the user was able to complete their task, either fully or partially, or not (see SM for details). Dropping cases with nonconforming labels left 78420 conversations for this analysis.

To test if users gain more satisfaction when they successfully complete more cognitively complex tasks, we used the following linear regression model which includes an interaction term between the level of cognitive complexity and task completion status. We recognize that conversation length, measured by the number of turns, can vary across cognitive complexity

$$Satisfaction \leftarrow Complexity * Completion + TurnNum * Completion + KnowledgeDomain$$

| | coef | std err | t | P> \|t\| | [0.025 | 0.975] |
|---|---|---|---|---|---|---|
| Understand | 3.9303 | 0.338 | 11.629 | 0.000 | 3.268 | 4.593 |
| Apply | -0.3979 | 0.322 | -1.235 | 0.217 | -1.030 | 0.234 |
| Analyze | 0.1581 | 0.429 | 0.368 | 0.713 | -0.683 | 0.999 |
| Evaluate | -0.2325 | 0.836 | -0.278 | 0.781 | -1.870 | 1.405 |
| Create | -7.3502 | 0.347 | -21.198 | 0.000 | -8.030 | -6.671 |
| Partially completed | 12.9145 | 0.212 | 60.937 | 0.000 | 12.499 | 13.330 |
| Completed | 16.6763 | 0.199 | 83.637 | 0.000 | 16.285 | 17.067 |
| Understand:Partially completed | 2.9378 | 0.405 | 7.256 | 0.000 | 2.144 | 3.731 |
| Apply:Partially completed | 4.6836 | 0.397 | 11.786 | 0.000 | 3.905 | 5.462 |
| Analyze:Partially completed | 4.5896 | 0.528 | 8.699 | 0.000 | 3.556 | 5.624 |
| Evaluate:Partially completed | 5.4953 | 1.077 | 5.101 | 0.000 | 3.384 | 7.607 |
| Create:Partially completed | 7.2740 | 0.444 | 16.388 | 0.000 | 6.404 | 8.144 |
| Understand:Completed | 0.7672 | 0.402 | 1.909 | 0.056 | -0.020 | 1.555 |
| Apply:Completed | 2.7471 | 0.389 | 7.066 | 0.000 | 1.985 | 3.509 |
| Analyze:Completed | 2.0954 | 0.562 | 3.728 | 0.000 | 0.994 | 3.197 |
| Evaluate:Completed | 2.0415 | 1.094 | 1.866 | 0.062 | -0.103 | 4.186 |
| Create:Completed | 8.2891 | 0.435 | 19.048 | 0.000 | 7.436 | 9.142 |
| NumUserMessages | 0.2678 | 0.039 | 6.791 | 0.000 | 0.191 | 0.345 |
| NumUserMessages:Partially completed | 0.2373 | 0.043 | 5.492 | 0.000 | 0.153 | 0.322 |
| NumUserMessages:Completed | 1.0933 | 0.048 | 22.587 | 0.000 | 0.998 | 1.188 |
| Intercept | -7.9653 | 0.159 | -50.020 | 0.000 | -8.277 | -7.653 |

**Table 1. Regression results where the dependent variable is user satisfaction.** In general, the more complex the task the more satisfied the user whether it was partially or totally completed.

categories and potentially influence users' satisfaction. Therefore, we incorporated an interaction between the number of turns and task completion to control for that confounding effect. Finally, we add a bias term to account for variance in average satisfaction across different knowledge domains. Table 1 shows results, with coefficients representing gains in user satisfaction for increasing levels of cognitive complexity compared to a baseline of the lowest complexity level (Remember). In general, for both partial and fully completed tasks, user satisfaction increases more as the level of task complexity increases. For example, for partially completed tasks, user satisfaction increases over baseline by 2.9 for a partially completed Understand task, an increase of roughly 4.6 for partially completed Apply and Analyze tasks, an increase of 5.5 for an Evaluate task and an increase of 7.9 for a partially completed Create task. The same trend largely holds for fully completed tasks with the exception that Evaluate had a non-significant coefficient due to there being very few Evaluate tasks (as shown in Figure 2).

## Knowledge Work and Task Complexity

Finally, we combine our analysis of knowledge work and task complexity at the conversation/session level to fully illustrate the qualitative shift in types of tasks being done in a generative search engine compared to a traditional one. Based on Figure 2, we binarized our task complexity rating as Remembering vs. not Remembering. Figure 3 shows that Bing Copilot conversations were more complex in nature, with roughly half of the domains containing majority complex conversations, whereas Bing Search had no domains that contained a majority complex sessions. Also, we rated each Bing Copilot conversation and Bing Search session for the degree to which the user task was knowledge work or not in nature. These ratings were done with GPT-4-based rating prompts, with human validation, (see SM) on Bing Search sessions and Bing Copilot conversations and were independent of one another and of the domain classification. Again, dropping cases with nonconforming labels from our LLM-based classifiers left 76541 Bing Copilot conversations and 78701 Bing Search sessions. Figure 3 largely confirms that the domains in Bing Copilot contain user tasks that skew more towards knowledge work. Prominent example domains include Data analysis and visualization (89% of conversations classified as knowledge work) Academic writing and editing (89%), and Web development (81%). In contrast, Bing Search saw only a single domain (Science, 64% of sessions) where more than half of search sessions were classified as knowledge work, and common Bing Search domains such as Social Media (1%) and Shopping (4%) included very few sessions classified as knowledge work.

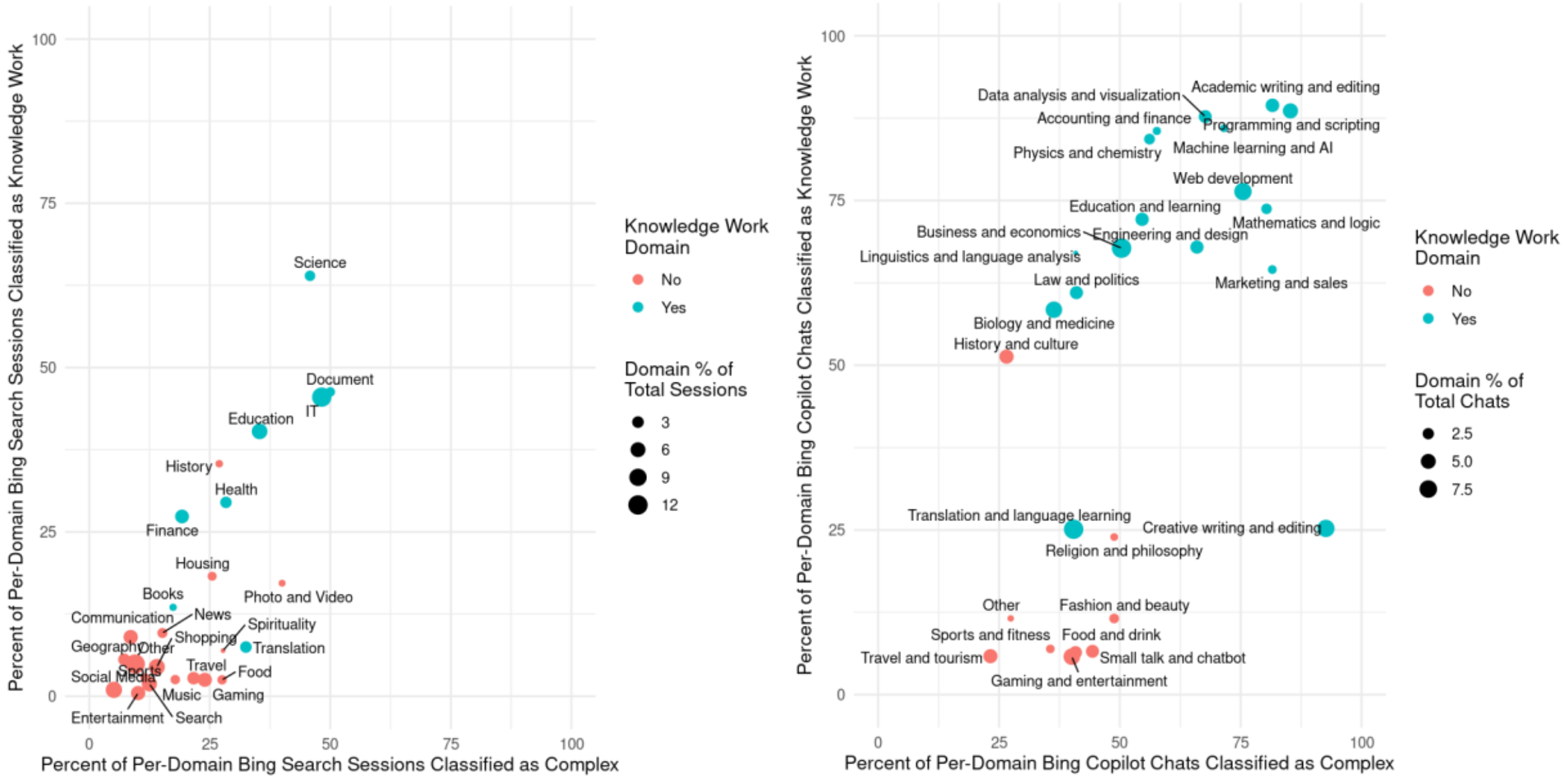

**Figure 3.** Within each domain, the x-axis is the fraction of sessions/conversations that are classified as complex vs. not and the y-axis is the fraction of sessions/conversations that are knowledge work vs not. The coloration of each domain is based on if it a knowledge work domain determined by alignment with U.S. BLS employment sectors. Observe that Bing Copilot has many more common domains in the high knowledge and high complexity quadrant than Bing Search.

Overall Figure 3 shows that all but one (Science) of the Bing Search domains reside in the minority knowledge work and minority complex quadrant whereas most of the domains, including many of the most common domains, of the Bing Copilot conversations reside in the majority knowledge work and majority complex quadrant. Furthermore, if we apply the same coding based on the alignment with BLS employment sectors as in Figure 1, we see that many of the domains coded as knowledge work do have a majority of conversations/sessions classified as knowledge work thereby corroborating conversation level findings with the suggestive domain level findings.

## Discussion

In summation, our work shows that people use generative search engines primarily for knowledge work tasks which are more complex than tasks commonly done with a traditional search engine. In addition, the more complex the task, the more satisfied the user was upon completing it.

When people use a generative search engine, they are interacting with online information differently than when they use a traditional search engine. They are not simply looking information up but instead doing tasks that involve processing that information, for example by analyzing or evaluating it. Since people are using generative search engines, which are in part based on LLMs, for more complex tasks in domains that were either not done at all, or done far less commonly, using traditional search engines, there is evidence that LLMs are helping people do tasks that they had to do in a much more manual fashion previously. This suggests that the rise of LLMs has shifted the frontier of tasks that machines can do to include tasks that humans used to have to do, mostly on their own, and furthermore that people are successfully using those new capabilities to their advantage. That people are doing more complex tasks frequently in the context of knowledge work shows that this type of work is being disrupted by LLMs.

More broadly, billions of people still use and will continue to use traditional search engines. Prior work has shown that the productivity gains from the information age did not achieve the growth rates that many expected and hoped for *(25)*. While the use of generative search engines is still growing, should they achieve the widespread adoption of their traditional counterparts we could potentially see a societal-scale shift in how people access online information and how they get work done. For example, many of the productivity boosts seen in the experimental setting *(4-8)* could translate to the societal scale.

The effects of the disruption of knowledge by LLMs might not all be positive. This work contributes real world empirical evidence of that shift to recent estimates of the impact of AI on job types *(26, 27)*. It is well known that automation has had a dramatic impact on the manufacturing sector in the United States *(28)* and it has been conjectured that knowledge work may be the next form of labor to be disrupted by AI. Our results indicate that knowledge work is already seeing a dramatic impact from LLMs. Further research is required to determine if LLMs are automating knowledge work, augmenting knowledge work, or if that is even the right framing of the question.

There are several limitations to this work. First, we have studied LLM usage for only two months, an early snapshot in time. Conversely, humanity has about 30 years of experience using traditional search engines. While it is important to understand the early trends, it is equally important to acknowledge they are early. We will need to see how trends evolve over time. Furthermore, advances in AI systems are occurring extremely rapidly. We will also need to monitor the effects of upcoming innovations. In general, we believe that understanding the effects of AI on society is one of the most important research questions facing humanity today so that we can try to nudge the impact of AI to be positive for as many people on the planet as possible.

**Acknowledgments:** This work was approved by Microsoft's IRB under record #10750.

**Author contributions:**

Conceptualization: SS, SC, LW, CC, RW, CS

Data Curation: SS, SC, LW, CC

Methodology: SS, SC, LW, CC, MW, JN, TS

Formal Analysis: SS, SC, LW, CC, TS

Visualization: SS, SC

Project administration: LY

Resources: RA, GB, SM, NR, LY

Supervision: SS, SC, LY

Writing – original draft: SS, SC

Writing – review & editing: SS, SC

**Competing interests:** All of the authors except LW, CC, and CS are full time employees of Microsoft Corporation. This work was done while LW and CC were interns at Microsoft and CS was visiting Microsoft.

**Data and materials availability:** Aggregated data to replicate the figures in this research will be made available upon publication. Raw, individual level data will not be made public to protect user privacy and due to legal restrictions.

Supplementary Materials for

# The Use of Generative Search Engines for Knowledge Work and Complex Tasks

Siddharth Suri, Scott Counts, Leijie Wang, Chacha Chen, Mengting Wan, Tara Safavi, Jennifer Neville, Chirag Shah, Ryen W. White, Reid Andersen, Georg Buscher, Sathish Manivannan, Nagu Rangan, Longqi Yang

Corresponding author: suri@microsoft.com, counts@microsoft.com

**The PDF file includes:**

Supplementary Text
Tables S1 to S3
Materials and Methods

## Supplementary Text

Domain Taxonomy Generation and Validation

Our taxonomies of Bing Copilot conversations and Bing Search sessions were done using the following procedure. Since this is a general procedure we outline it here and provide our robustness checks and validation on our data sets. In addition, we refer the reader to *(11)* for more general validation and additional robustness checks.

1. Randomly sample 2000 conversations/sessions from the first week of data
2. Summarize each conversation/session using GPT-4 into descriptions of no more than 10 words.
3. Pass the summaries, in batches of 100, along with the current taxonomy, which is initially empty, into a prompt that develops a taxonomy of certain size, in our case 25, that encapsulates each batch. If the current taxonomy was insufficient to label a new batch of summaries, the model was instructed to create a new taxonomy that could label all summaries, including the new batch.
4. Repeat step 3 for each batch. In our case that means 20 times (once for each batch of size 100).

For the exact prompts used for these steps see Prompt: Iterative Label Generation in this SM. In addition, these labels and definitions were used in a new prompt to categorize all Bing Copilot conversations in our primary dataset (see Prompt: Domains).

*Robustness*

Step 2 and the subsequent operation on short summaries of the conversations is required because operating on the raw summaries requires too many tokens for GPT-4 to analyze. Summarizing also reduces noise as it allows the taxonomy to focus on the main task in the conversation.

To check the robustness of our domain generation algorithm we repeated it six times. Instead of sampling over the first eight weeks in step 1, we pulled three random samples from across the entire eight weeks. We used each of these three samples twice, randomizing the assignment of conversation summaries to the batches of 100 that were fed into the taxonomy creation prompt. Of the 16 most popular domains, making up 86.4% percent of total conversation sample, allowing for reasonable “soft matches” in domain labels to account for the non-deterministic nature of language models, 14 of the domains appeared in at least 4 of the 6 alternative taxonomies and sometimes in all 6. A soft match would include Engineering and Technology matching the Engineering and Design domain but would not include Mathematics and Statistics as a match for Data Analysis and Visualization. The two exceptions were Small Talk and chatbot (3.7% of conversations) and Data Analysis and Visualization (3.5% of conversations). Thus, we conclude that, at least for the most common domains, our taxonomy is robust.

*Validation*

Overall, we validated our Bing Search and Bing Copilot domain taxonomies by manually checking the accuracy of 250 Bing Search sessions and another 250 Bing Copilot conversations for a total of 500 manual checks of classifications of user input.

To validate the LLM’s classification of our Bing Search domains we manually checked 250 classifications for correctness. More specifically, we took a random sample of 10 search sessions from each of the 24 of the 25 domains and manually checked that their classification was correct. For the “Other” domain we asked if each search session would have been more correctly classified by a different domain. We found that only 3 out of the 10 search sessions in Other would have been more correctly classified in a different domain. Search sessions not in English were translated to English using an online translator. When the topic of the search was unfamiliar to the rater, we manually performed the search using a different search engine. These domains are ordered from the most to least common. Note that all of the top 10 domains (excluding Other) had either 9/10 or 10/10 correctly labelled search sessions (See Table S1). With the exception of Translation and Other every domain had at least 8/10 correct. Translation had a low correctness because the LLM would often categorize queries in different languages as Translation queries even if the person searching did not indicate that they were looking for a translation of their query. We also note that the domains were very well separated, with the exception of Travel and Geography, as it was easy to confirm that each search session in a specific domain did not fit better in a different domain.

**Table S1:** Bing Search domain validation

| Bing Search Domain | Num Correct | Num Incorrect | Knowledge Work? |
|---|---|---|---|
| Other | 7 | 3 | |
| IT | 10 | | Yes |
| Education | 9 | 1 | Yes |
| Social media | 10 | | No |
| Shopping | 10 | | No |
| Search | 10 | | No |
| Communication | 10 | | No |
| Entertainment | 10 | | No |
| Gaming | 10 | | No |
| Finance | 9 | 1 | Yes |
| Travel | 10 | | No |
| Translation | 3 | 7 | No |
| Geography | 10 | | No |
| Health | 10 | | Yes |
| Science | 9 | 1 | Yes |
| News | 8 | 2 | No |
| Document | 8 | 2 | Yes |
| Sports | 10 | | No |
| Food | 10 | | No |
| Music | 10 | | No |
| Housing | 8 | 2 | No |
| History | 10 | | No |
| Books | 9 | 1 | Yes |
| Photo and Video | 8 | 2 | No |
| Spirituality | 9 | 1 | No |

To validate the LLM's classification of our Bing Copilot domains we manually checked 250 classifications for correctness. More specifically, we took a random sample of 10 Bing Copilot conversations from each of the 25 domains and manually checked that their classification was correct. We considered an answer to be wrong if it was better served by a different label or if the label given was not appropriate for the conversation. For the "Other" domain we asked if each Bing Copilot conversation would have been more correctly classified by a different domain. For 21 out of the 25 domains had at least 8 out of 10 classifications correct (See Table S2). Often, when a classification was deemed incorrect it maintained a commonality with the labeled domain. For example, one conversation incorrectly labelled Creative Writing the user asked for a photo realistic, high-resolution card for a baby. This is a creative task but not one involving much writing. Translation had a lower correctness because the LLM would often categorize short conversations, and therefore containing little context, in different languages as Translation queries even if the person did not indicate that the conversation was specifically about translating something.

**Table S2:** Bing Copilot domain validation

| Bing Copilot Domain | Num Correct | Num Incorrect | Knowledge Work? |
|---|---|---|---|
| Business and Economics | 9 | 1 | Yes |
| Translation and Language Learning | 6 | 4 | Yes |
| Web Development | 8 | 2 | Yes |
| Creative Writing and Editing | 8 | 2 | Yes |
| Gaming and Entertainment | 8 | 2 | No |
| Biology and medicine | 10 | | Yes |
| Programming and scripting | 8 | 2 | Yes |
| History and culture | 10 | | No |
| Travel and tourism | 10 | | No |
| Academic writing and editing | 10 | | Yes |
| Education and learning | 10 | | Yes |
| Smalltalk and Copilotbot | 8 | 2 | No |
| Engineering and design | 9 | 1 | Yes |
| Law and politics | 10 | | Yes |
| Data analysis and visualization | 7 | 3 | Yes |
| Food and drink | 9 | 1 | No |
| Physics and chemistry | 10 | | Yes |
| Mathematics and logic | 10 | | Yes |
| Fashion and beauty | 7 | 3 | No |
| Marketing and sales | 10 | | Yes |
| Sports and fitness | 10 | | No |
| Accounting and finance | 10 | | Yes |
| Machine learning and AI | 9 | 1 | Yes |
| Religion and philosophy | 9 | 1 | No |
| Other | 6 | 4 | |
| Linguistics and language analysis | 10 | | Yes |

Knowledge Work Domain Coding Validation

The domains used in this work were created bottom-up from Bing Copilot conversation text. Thus, we did not assume any *a priori* alignment with any existing taxonomies of work tasks. We validate our labeling of domains as knowledge work domains in two ways. First, we make a best effort alignment with North American Industry Classification System (NAICS) codes used for industry classification by the United States Bureau of Labor Statistics. All domains considered knowledge work domains fall into one of four sectors that we consider reasonable proxies for k related work: Professional, Scientific, and Technical Services sector (2-digit NAICS code 54), Information (51), Finance and Insurance (52), and Educational Services (61).

The conversation level classifications shown in Figure 2 help validate the coding of domains into knowledge work versus not knowledge work. For Bing Copilot, only two domains (Creative Writing and Editing; Translation and Language Learning) coded as knowledge work contain less than 50% of conversations themselves classified directly as knowledge work, and both still contained more than a third of conversations classified as knowledge work. The remaining domains generally are well above a knowledge work conversation majority.

Task Complexity Validation

Our task complexity prompt is a faithful implementation of taxonomy Anderson and Krathwohl's. To validate this prompt the first two authors manually classified a random sample of 100 Bing Copilot conversations according to the six elements of the taxonomy. To measure their agreement with the prompt output we used Cohen's kappa with quadratic weights. Their average interrater reliability with the prompt output was 0.76 indicating high agreement.

User Satisfaction and Task Completion Validation

We refer the reader to *(24)* for extensive validation and robustness checks as to how we measure a user's satisfaction with their Bing Copilot conversation.

We validated our measure of task completion in two ways. First, we manually labelled 99 random English Bing Copilot conversations as either not completed, partially completed or fully completed. We then looked at a binarized metric where we combined partially completed and fully completed, in this case the model agreed with our human generated categorization in 85 out of 99 conversations. In the ternary metric which separates not completed, partially completed or fully completed our model agreed with our human categorizations in 62 out of 99 conversations indicating that the majority of the discrepancies (23 out of 37) were due to the model not being able to differentiate between a partially vs fully completed task.

Robustness of Satisfaction / Completion Regression Analysis

A small percent (approximately 3.5% of the full dataset) or Bing Copilot conversations did not contain an AI response and instead contained only a single utterance from the user. This could be for various reasons, including the user closing the session prior to the AI response or the user manually stopping the AI response. These conversations were labeled 'Not completed', and while such a classification is arguably accurate given the conversation log data available, we reran the regression of task complexity by task completion on user satisfaction after filtering out conversations with no AI response. The results, as shown in Table S3, remain substantively the same, with the same pattern of generally larger coefficients on increasing levels of task complexity, conditional on the user at least partially completing their task.

**Table S3:** Regression of task complexity by task completion on user satisfaction while excluding incomplete conversations.

| | coef | std err | t | P> \|t\| | [0.025 | 0.975] |
|---|---|---|---|---|---|---|
| **Understand** | 4.9822 | 0.361 | 13.809 | 0.000 | 4.275 | 5.689 |
| **Apply** | 0.6275 | 0.346 | 1.816 | 0.069 | -0.050 | 1.305 |
| **Analyze** | 1.1938 | 0.448 | 2.662 | 0.008 | 0.315 | 2.073 |
| **Evaluate** | 1.1112 | 0.880 | 1.263 | 0.206 | -0.613 | 2.835 |
| **Create** | -6.3402 | 0.371 | -17.092 | 0.000 | -7.067 | -5.613 |
| **Partially completed** | 14.1272 | 0.228 | 61.923 | 0.000 | 13.680 | 14.574 |
| **Completed** | 17.9081 | 0.216 | 82.719 | 0.000 | 17.484 | 18.332 |
| **Understand:Partially completed** | 1.8876 | 0.425 | 4.446 | 0.000 | 1.055 | 2.720 |
| **Apply:Partially completed** | 3.6613 | 0.417 | 8.778 | 0.000 | 2.844 | 4.479 |
| **Analyze:Partially completed** | 3.5617 | 0.544 | 6.545 | 0.000 | 2.495 | 4.628 |
| **Evaluate:Partially completed** | 4.1773 | 1.114 | 3.750 | 0.000 | 1.994 | 6.361 |
| **Create:Partially completed** | 6.2838 | 0.464 | 13.546 | 0.000 | 5.375 | 7.193 |
| **Understand:Completed** | -0.2977 | 0.422 | -0.706 | 0.480 | -1.124 | 0.529 |
| **Apply:Completed** | 1.7140 | 0.409 | 4.192 | 0.000 | 0.913 | 2.515 |
| **Analyze:Completed** | 1.1144 | 0.578 | 1.926 | 0.054 | -0.019 | 2.248 |
| **Evaluate:Completed** | 0.7109 | 1.130 | 0.629 | 0.529 | -1.505 | 2.927 |
| **Create:Completed** | 7.2672 | 0.455 | 15.960 | 0.000 | 6.375 | 8.160 |
| **NumUserMessages** | 0.3305 | 0.040 | 8.229 | 0.000 | 0.252 | 0.409 |
| **NumUserMessages:Partially completed** | 0.1738 | 0.044 | 3.960 | 0.000 | 0.088 | 0.260 |
| **NumUserMessages:Completed** | 1.0284 | 0.049 | 20.962 | 0.000 | 0.932 | 1.125 |
| **Intercept** | -9.1754 | 0.180 | -51.028 | 0.000 | -9.528 | -8.823 |

## Materials and Methods

Prompts

We provide here all prompts used for taxonomy generation and classification, all of which were run using GPT-4. We note that the generative nature of prompts means that the output (specified to be in XML format) is not always perfectly formatted. As indicated in the main text, we discarded any conversation that did not receive clean classification labels from all classifications as output from the prompts. This resulted in the elimination of 3.3% of Bing Copilot conversations and 1.6% of Bing Search sessions from the final datasets.

We note that with the exception of the domains prompts, we provide the prompt for Bing Copilot only, as the equivalents for Bing Search were near verbatim similar, with only the minimum changes needed (e.g., replacing the words "conversation history" with "search session history").

*Iterative Label Generation*

<|instruction|>

# Context and data description:

- Your primary goal is to review a given intent taxonomy and make adjustments if needed.

- The given taxonomy is in the **yaml** format, where each entity in this taxonomy is a domain category. You can use this taxonomy to help you construct a new domain taxonomy. The schema of this domain taxonomy is as follows:

  - **title**: the title of the domain category
  - **description**: the description of the domain category
  - **examples**: a list of examples in the domain category

Here is an example of a taxonomy:

{insert example of a taxonomy here}

# Your primary goal is to generate a taxonomy that can serve for the following use cases:

- The primary use case of this taxonomy is to help understand user's topical domains in human-AI conversations. Entities in this taxonomy can be used to label **user topical domains** in human-AI conversations.

# Here includes some criteria of a generic taxonomy:

- **Accuracy**: The definitions, descriptions of classes, properties, and individuals in a taxonomy should be correct.

- **Completeness**: All the data should be reliably classified using this taxonomy.

- **Conciseness**: The taxonomy should not include any irrelevant elements with regards to the user topical domain in AI Copilot.

- **Clarity**: The taxonomy should communicate the intended meaning of the defined terms. Definitions should be objective and independent of the context.

- **Consistency**: The taxonomy does not include or allow for any contradictions.

# Requirements of your output taxonomy:

- Your output **topical domain** taxonomy should focus on the user's task objects. This is **different** from an **intent** taxonomy, which primarily describes the task actions.

- Your output taxonomy should match the existing taxonomy and the data as closely as possible, without leaving out important topical domain categories or including unnecessary ones. Please make sure there is no overlap or contradiction among the topical domain categories in your output taxonomy.

- Your output *title* of each category should be *no more than 5 words*. The title should be a concise and clear label for the topical domain category.

- Your output *description* of each category should be *no more than 30 words*. The description should explain the user's nature of the topical domain category and should differentiate it from other topical domain categories.
- The number of examples for each topical domain category should be *no more than 3*. The examples should either come from the given taxonomy or the provided data with **exactly the same content**. Please do not invent new examples or topical domains that are not in the given taxonomy or the data.
- **Size limit of the output taxonomy**: The total number of topical domain categories should be **no more than 50**.
- Your output taxonomy and examples should be in *English* only.
# You are asked to answer the following questions:
- Q1. Please check the above general criteria and the specific taxonomy requirements one-by-one. Does the provided taxonomy satisfy the above requirements, word limits and taxonomy size limit? Please answer "yes" or "no". If there is no given taxonomy, please answer "no".
- Q2. Please explain your answer to Q1. If your answer to Q1 is "no", please also describe if you'd like to construct the taxonomy structure from scratch or you plan to make changes on the given taxonomy. Your answer to this question should be **within 100 words**.
- Q3. If your answer to Q1 is "no", then generate a new domain taxonomy from the given data and the given existing taxonomy (if available). Your output taxonomy should be in the **table** format with the same schema. If your answer to Q1 is "yes", please answer "N/A". Please make sure the new taxonomy satisfies **all of the above requirements**. Please **do not** invent new examples or new topical domains that are not in the existing taxonomy or the provided data.

# Tips:
- If you're given an existing taxonomy, you can use the provided data to update this taxonomy. By incorporating the newly provided data, you can *add new categories*, *merge or generalize existing categories*, *split existing categories*, *reorganize the current tree structure*, *change titles and descriptions*, *swap examples* and do other operations if needed.
- If the topical domain category structure of the given taxonomy cannot be easily adjusted, then please construct a new structure of these topical domain categories based on their descriptions and the provided examples. Please make sure your new taxonomy covers the semantics of the existing taxonomy as thoroughly as possible. Please **do not** invent new topical domains that are not in the existing taxonomy or the provided data.
- You should carefully review the provided examples in each category and make sure they are correctly labeled. You can also reorganize the examples or create new categories from them when needed. You're allowed to have fewer than 3 for each category but your examples should only come from examples in **the given taxonomy or the provided data**. Please **do not** invent new examples that are not in the existing taxonomy or the provided data.
- Please make sure your new taxonomy satisfies the **word limits** and **taxonomy size limit**. You're allowed to have fewer than 50 categories in your final output. If you couldn't fit your new taxonomy into the limits, please consider merging or abstracting some specific categories into more general categories.
- Please make sure there is no overlaps or contradictions among the topical domain categories in your output taxonomy.

# Please provide your answers in the following format:

Please provide your answers between the tags: <Q1>your answer to Q1</Q1>, <Q2>your answer to Q2</Q2> and <Q3>your final output taxonomy</Q3>. Please make sure your answers are within the word limits.

<|end of instruction|>

<|provided data|>

{new data here}

<|end of data|>

<|existing taxonomy|>

{insert taxonomy here}

<|end of taxonomy|>

<|end of prompt|>

*Domains– Bing Copilot*

<|Instruction|>

You will be given a conversation history between a User and an AI agent. Your task is to label the topical domain of the conversation using one of the topical domain categories listed below.

The categories and definitions of topical domains are:

- Web development: The user's topical domain is related to creating, modifying or debugging websites or web applications using various languages, frameworks or tools.
- Programming and scripting: The user's topical domain is related to creating, modifying or debugging computer programs or scripts using various languages, frameworks or tools.
- Data analysis and visualization: The user's topical domain is related to collecting, analyzing, visualizing or presenting data using various methods, tools or software.
- Machine learning and AI: The user's topical domain is related to applying or developing artificial intelligence or machine learning techniques or models using various methods, tools or software.
- Engineering and design: The user's topical domain is related to the application of engineering principles or methods to various fields, such as civil, mechanical, electrical, chemical, etc., or to the design or creation of visual or auditory works.
- Academic writing and editing: The user's topical domain is related to the creation, improvement or evaluation of texts for academic purposes or audiences, such as essays, articles, theses, etc., or to the citation, referencing or plagiarism issues of academic writing.
- Creative writing and editing: The user's topical domain is related to the creation, improvement or evaluation of texts for creative or imaginative purposes or audiences, such as stories, poems, songs, etc., or to the genres, styles or themes of creative writing.
- Translation and language learning: The user's topical domain is related to the translation of words, phrases or sentences in different languages, or to the learning or practice of languages or linguistic features, rules or systems.
- Linguistics and language analysis: The user's topical domain is related to the study or analysis of the structure, function, evolution or diversity of languages or linguistic features, rules or systems.
- Physics and chemistry: The user's topical domain is related to the study of the matter, energy, forces, reactions or interactions of the physical or chemical world, or to the concepts, theories or experiments of various branches of physics or chemistry.

- Biology and medicine: The user's topical domain is related to the study of the structure, function, evolution or diversity of living organisms, or to the diagnosis, treatment or prevention of diseases or disorders, or to the physical or mental well-being of humans or animals.
- Mathematics and logic: The user's topical domain is related to the application or solution of mathematical or logical problems or puzzles, such as arithmetic, algebra, geometry, calculus, etc., or to the concepts, theories or arguments of various branches of mathematics or logic.
- Business and economics: The user's topical domain is related to the operation, strategy or analysis of an organization or an industry, or to the study of the production, distribution or consumption of goods or services, or to the market, trade or currency issues.
- Accounting and finance: The user's topical domain is related to the recording, reporting or auditing of financial transactions or statements, or to the management, investment or valuation of money or assets, or to the financial products, services or regulations.
- Marketing and sales: The user's topical domain is related to promoting or selling products or services, or to creating, evaluating or improving advertisements or marketing strategies or campaigns, or to persuading or influencing potential customers or clients.
- Education and learning: The user's topical domain is related to learning, teaching or studying various subjects or skills, or to the methods, tools or resources for education or learning, or to the educational institutions, programs or policies.
- Gaming and entertainment: The user's topical domain is related to the playing, designing or reviewing of video games or other interactive media, or to the enjoyment, amusement or diversion of oneself or others, such as movies, shows, books, jokes, etc.
- Sports and fitness: The user's topical domain is related to the physical activities, exercises or games that involve skill, strength or endurance, or to the equipment, rules or strategies of sports or fitness, or to the sports teams, players or events.
- History and culture: The user's topical domain is related to the past or present events, people, places, customs, beliefs, values or arts of a group of people or a region, or to the appreciation, criticism or creation of historical or cultural works.
- Law and politics: The user's topical domain is related to the rules, regulations or principles that govern the conduct or relations of people or entities, or to the rights, responsibilities or remedies of legal subjects or cases, or to the opinions, issues or actions of political actors or institutions.
- Religion and philosophy: The user's topical domain is related to the belief, worship or practice of a supernatural or divine power or entity, or to the personal or philosophical quest for meaning or purpose, or to the concepts, theories or arguments of various schools of thought.
- Fashion and beauty: The user's topical domain is related to the style, appearance or attractiveness of clothing, accessories, cosmetics or hair, or to the trends, tips or advice of fashion or beauty, or to the fashion or beauty products, services or brands.
- Food and drink: The user's topical domain is related to the preparation, storage or consumption of food or beverages, or to the recipes, ingredients, flavors or health benefits of food or beverages, or to the preferences, opinions or recommendations of food or beverages.
- Travel and tourism: The user's topical domain is related to the movement, exploration or discovery of different places or cultures, or to the transportation, accommodation or activities of travelers, or to the travel guides, tips or advice.
- Small talk and Copilotbot: The user's topical domain is related to the casual, friendly or polite conversation or interaction with the AI agent, or to the general questions or answers about the AI agent or various topics or domains.
- Other: the user's topical domain does not fit into any of the above specified topical domains.

# INTENT OBJECTS

An intent object is defined as the thing that the user is trying to find, analyze, create, or achieve. There are no predefined categories for intent objects.

# TASK

Q1. Classify the user's primary topical domain in the conversation. Your answer to this for each conversation should be a **single** domain. Your answers should be in the provided list of topical domain categories.

Q2. Provide a brief explanation of your answer to Q3.

Q3. Generate a description in 10 words or fewer of the user's main intent object in the conversation. Your answer to this for each conversation should be a **single** description in 10 words or fewer.

# ANSWER FORMAT

Provide your answers in XML format between the tags <Q1>{your answer to the topical domain in Q1}</Q1>, <Q2>{your answer to Q2}</Q2>, <Q3>{your answer to the intent object in Q3}</Q3>.

# TIPS

- Provide all answers in **English**.
- Some users may switch between languages during the conversation. The conversation is **not** necessarily Creation intent unless the user **explicitly asks the agent to translate a piece of text**.

<|end of instruction|>

<|Conversation between User and AI|>

Conversation:

{insert conversation here}

<|end of Conversation|>

<|end of prompt|>

*Domains – Bing Search*

<|Instruction|>

# OVERVIEW

You will be given search session queries and a topical domain taxonomy. Your task is to classify the search session based on the given topical domain taxonomy.

# Your goal is to answer the following questions:

Q1. Classify the user's topical domain in the search session. If **multiple** domains are detected, provide **the most prominent** domain.

Q2. Provide a brief explanation of your answer to Q2.

# Requirements of your output:

# Provide your answers in xml format between the tags <Q1>{your answer to Q1}</Q1>, <Q2>{your answer to Q2}</Q2>.

# Please make sure your answers are in **English** and in the above given xml format.

## TOPICAL DOMAIN TAXONOMY

|title|description|examples|

|Search|Online tools for web search.|Bing AI; Baidu; Google|

|Communication|Online tools for text, voice, or video communication.|WhatsApp Web; Gmail; Outlook|

|Social Media|Online platforms for social media and video content.|Facebook; YouTube; Instagram|

|Translation|Online tools for text, speech, or image translation.|Google Translate; Baidu translation; Reverso|

|News|Online platforms for news and current affairs.|Fox News; BBC news; @ hfxgov|

|Music|Online tools for music streaming, downloading, or creation.|Spotify; SoundCloud; Music streaming quality|

|Education|Online platforms for education and learning content, courses, or tools.|Geogebra; AI class; Online courses in Tamil language|

|Science|Online platforms for science information, research, or education.|NASA; Microbial genome size data; MT7622 chip|

|IT|Online or offline tools for IT-related queries, solutions, or products.|Windows 11 troubleshooting; reset local password windows 10; How to transfer files to Android subsystem|

|Document|Online or offline tools for document creation, editing, conversion, or sharing.|PDF compression; Hancom Docs; Ali cloud disk|

|Shopping|Online or offline platforms for buying or selling goods or services.|Ebay; Amazon; Tractor parts from Kumar Bros USA|

|Travel|Online or offline platforms for travel information, booking, navigation, or reviews.|Booking.com; Skyscanner; Ko Panyi, Thailand|

|Health|Online or offline platforms for health information, advice, products, or services.|Skin anatomy; Osteoporosis; Health insurance in Maryland|

|Gaming|Online or offline platforms for gaming information, streaming, downloading, or playing.|Minecraft Dungeons Arch-Illager; mod manager; Plants vs Zombies video game|

|Photo and Video|Online or offline tools for photo and video information, editing, sharing, or products.|Video editing software for Android; Wildlife videos from Gold Harbour; Wallpapers of Jennie and Rose from Blackpink|

|Entertainment|Online or offline platforms for entertainment information, streaming, downloading, or sharing.|Netflix; Hulu; Xu Kaicheng|

|Dating|Online or offline platforms for dating information or services.|Online dating and video streaming; I want a boyfriend; Badoo|

|Spirituality|Online or offline platforms for spirituality information, services, or products.|Buddhism and meditation; Astrology and horoscopes; Raef login support|

|Books|Online or offline platforms for books information, reading, downloading, or sharing.|PDF books; Kindle; Romance novel about ex-convict wife|

|Food|Online or offline platforms for food, recipes, or cooking information, services, or products.|Beet and arugula salad recipe; Green tea and fat burning; Rhubarb and strawberry pie|

|Sports|Online or offline platforms for sports information, services, or products.|Baseball; Cribbage online; Beekeeping in Oulu|

|Finance|Online or offline platforms for finance information, services, or products.|Wells Fargo login; Stock market color codes; ADP|

|Housing|Online or offline platforms for housing information, services, or products.|House search; Mesa landforms; avra imperial hotel|

|History|Online or offline platforms for history information, services, or products.|Sununu; Bievre river and its history; Xi Jinping's informationization and ideological implementation|

|Geography|Online or offline platforms for geography information, services, or products.|Texel (Dutch island); Metro station and map of Czech Republic; Tschanüff Castle in Valsot, Switzerland|

|Other|The user's topical domain does not fit into any of the above specified topical domains.||
<|end of instruction|>

<|search session|>
Queries in this search session:
{insert search sessions here}
<|end of search session|>
<|end of prompt|>

Knowledge Work
<|Instruction|>
# OVERVIEW
You will be given a conversation history between a User and an AI agent. Your task is to determine whether the user's task is related to knowledge work or not.
-Knowlege work: Knowledge work concerns the creation, handling, and distribution of information and knowledge products. Knowledge work involves non-routine tasks and relies on the ability to think creatively and analytically in a way that combines convergent and divergent thinking. Examples include creating strategic business plans, designing software, and conducting scientific research.
-Not knowledge work: The user's conversation is not related to knowledge work. In these cases, the conversation may be about manual work or may be related to non-work topics like entertainment, humor, or general topics like religion and politics.
- Both: The user's conversation is related to a mix of knowledge work and not knowledge work.
-None: The user's conversation is not at all related to knowledge work or not knowledge work or a mix of the two.

# TASK
Q1. Classify the whether the user is engaged in a task that is related to knowledge work or not. Your answer to this for each conversation should be a **single** category label. Your answer should be from one of four categories listed above: Knowledge work, Not knowledge work, Both, None.
Q2. Provide a brief explanation of your answer to Q1.
# ANSWER FORMAT
Provide your answers in XML format between the tags <Q1>{your answer to Q1}</Q1>, <Q2>{your answer to Q2}</Q2>.

# TIPS
- Provide all answers in **English**.
- Some users may switch between languages during the conversation.
<|end of instruction|>

<|Conversation between User and AI|>
Conversation:
{insert conversation here}
<|end of Conversation|>
<|end of prompt|>

*Task Complexity*

# OVERVIEW

You will be given a conversation history between a User and an AI agent. Your task is to answer questions about the complexity of the user's primary task.

# TASK COMPLEXITY

Task complexity is characterized by the level of cognitive processes and knowledge required to successfully complete a task by human beings. Specifically task complexity is defined using the following 6-type taxonomy of learning objectives:

- Task complexity and learning objectives taxonomy type 1, Remember: These tasks involve retrieving, recalling, and recognizing relevant knowledge from long-term memory or from information archives such as search engines, webpages, wikis, or libraries. Often the user wants to find factual information or answers to specific questions. Recalling involves retrieving relevant knowledge from long-term memory. Recognizing involves retrieving relevant knowledge from long-term memory in order to compare it with presented information.
- Task complexity and learning objectives taxonomy type 2, Understand: Understanding takes place when a person is building a connection between knowledge that is new to them and their prior knowledge. This can take many possible forms:
    - Interpreting: Interpreting occurs when a person converts information from one form or representation to another, for example, from words to numbers or numbers to words, or words to an image, or an image to words.
    - Exemplifying: Exemplifying occurs when someone gives a specific example or instance of a general concept or principle.
    - Classifying: Classifying occurs when someone recognizes that something (e.g., a particular instance or example) belongs to a certain category
    - Summarizing: Summarizing occurs when someone wants a short statement that represents presented information ar abstracts a general theme.
    - Inferring: Inferring involves finding a pattem within a series of examples or instances.
    - Comparing: Comparing involves detecting similarities and differences between two or more objects, events, ideas, problems, or situations. Comparing includes finding one-to-one correspondences between elements and patterns in one object, event, or idea and those in another object, event, or idea.
    - Explaining: Explaining occurs when someone is able to construct and use a cause-and-effect model of a system and use the model to determine how a change in one part of the system or one "link" in the chain affects a change in another part.
- Task complexity and learning objectives taxonomy type 3, Apply: Applying involves using procedures to perform exercises or solve problems. Thist could take two forms.
    - Executing: It could be that the user know the procedure to be done and is asking the AI to do it.
    - Implementing: The suer does not know the procedre to be done but is asking the AI to do it.
- Task complexity and learning objectives taxonomy type 4, Analyze: Analyze involves breaking material into its constituent parts and determining how the parts are related to one another and to an overall structure. This could take several forms:
    - Differentiating: Differentiating involves distinguishing the parts of a whole structure in terms of their relevance or importance.

- Organizing: Organizing involves identifying the elements of a communication or situation and recognizing how they fit together into a coherent structure.
- Attributing: Attributing occurs when someone is able to ascertain the point of view, biases, values, or intention underlying communications
- Task complexity and learning objectives taxonomy type 5, Evaluate: Evaluate is defined as making judgments based on criteria and standards. This could take two forms:
- Checking: Checking involves testing for internal inconsistencies or fallacies in an operation or a product.
- Critiquing: Critiquing involves judging a product or operation based on extemally imposed criteria and standards.
- Task complexity and learning objectives taxonomy type 6, Create: Create involves putting elements together to form a coherent or functional whole.
- Other: If the task complexity and learning object could not be classified as any of the previous types, label it "Other".

# Task:
Q1. Classify the user's primary task in the conversation with the AI agent into one, and only one, of the six task complexity categories: Remember, Understand, Apply, Analyze, Evaluate, or Creating. ** Only ** provide the category name in your answer.
Q2. Provide a brief explanation for your answer to Q1: why did you classify the user's conversation with the AI agent into the task complexity category that you selected? Your explanation should be no more than 20 words.

# ANSWER FORMAT
Provide your answers in XML format between the tags <Q1>{your answer to the Q1}</Q1><Q2>{your answer to the Q2}</Q2>

<|conversation|>
{insert conversation here}
<|endofconversation|>
<|endofprompt|>

Task Completion
<|Instruction|>
# You will be given a conversation history between a User and an AI agent. You will be asked to answer questions related to the user task in this conversation.
# You are asked to answer these questions in **English**
## Q1. Please summarize the main user task and classify the main domain of this conversation. A user task should start with a verb.
## Q2. Explain your answer to Q1 **within 30 words**.
## Q3. Please classify if the user task is completed in this conversation with the assistance of the AI into the following categories:
- Not completed
- Partially completed
- Appears to be completed: the task appears to be completed but lacks user confirmation
- Explicitly completed: the task is fully completed and with user confirmation.
## Q4. Explain your answer to Q3 **within 30 words**.

## Q5. Based on your answers to Q3 and Q4 and the conversation content, please provide a task completion rating score on a scale of 1 (Not completed) to 10 (Explicitly completed).

## Q6. Explain your answer to Q5 **within 50 words**.

# Provide your answers in xml format between the tags <Q1.MainTask>{your answer to the main user task in Q1}</Q1.MainTask>, <Q1.MainDomain>{your answer to the main task domain in Q1}</Q1.MainDomain>, <Q2>{your answer to Q2}</Q2>, <Q3>{your answer to Q3}</Q3>, <Q4>{your answer to Q4}</Q4>, <Q5>{your answer to the 1-10 task completion rating in Q5}</Q5>, <Q6>{your answer to Q6}</Q6>.

# Please make sure your answers are in **English** and in the above given xml format.

<|end of instruction|>

<|Conversation between User and AI|>

Conversation:

{insert conversation here}

<|end of Conversation|>

<|end of prompt|>

User Satisfaction

# Your task is to evaluate both user satisfaction and dissatifaction with a conversational AI agent by applying the given rubrics to the given conversation history between the user and the agent.

# Rubric instructions

- Each rubric contains 10 criteria.
- Each criterion has a Yes or No statement.
- Your job is to go through the conversation history carefully and answer Y to each statement that applies to the conversation, then give the statement a score of 1-10 to reflect how likely the expressed sentiment will impact the user's overall satisfaction/dissatifaction with the interaction. If the statement is not applicable answer N and give an overall score of 0.
- The rubric is formatted in a table format with 10 rows and two columns: Index|Y/N Question

# SATISFACTION RUBRIC

1|The user thanks or compliments the AI agent for its responses.

2|The user continues to ask follow-up questions on the same or different topics.

3|The user does not express any confusion, frustration, or dissatisfaction with the AI agent's answers.

4|The user learns something new or useful by indicating curiosity and satisfaction with the information provided.

5|The user follows the AI agent's suggestions or instructions when applicable.

6|The user uses positive feedback words (e.g., excellent, amazing) or emojis, indicating enthusiasm and enjoyment of the conversation.

7|The user shares personal details or opinions with the AI agent when appropriate.

8|The user jokes with or challenges the AI agent in a friendly manner when suitable.

9|The user acknowledges or confirms that they understood or agreed with the AI agent's explanations when relevant.

10|The user ends the conversation on a positive note without asking for more information or assistance.

# DISSATISFACTION RUBRIC
1|The user explicitly expresses dissatisfaction, frustration, annoyance, or anger with the AI agent's response or behavior.
2|The user repeatedly asks the same or similar questions or requests.
3|The user corrects the AI agent's mistakes or inaccuracies in its information or output.
4|The user changes the topic abruptly without acknowledging or following up on the previous one(s).
5|The user does not respond to the AI agent's questions, suggestions, feedback requests, etc.
6|The user has unrealistic expectations of what the AI agent can do and does not accept its limitations or alternatives.
7|The user implies that their query was ignored completely or that the response did not address their intent/goal at all.
8|The user perceives a decline in quality of service compared to previous experience with other agents/tools, etc.
9|The user wants more specific/useful information than what is provided by the AI agent.
10|The user feels that there is a mismatch between their preferred style and what is provided by the AI agent.

# Task:
- Go through the conversation history thoroughly
- For each rubric question think about your answer to each question carefully.
- Answer Y or N only to each rubric question.
- For Y answer, score your answer on a scale of 1-10 (low to high) to reflect how likely the expressed sentiment will impact the user's overall satisfaction/dissatisfaction with the interaction. For N answer, score 0.
- Only provide ONE most confident answer to each question.
- You *MUST* output your answers to all 10 questions provided in each rubric.

# Conversation:
<insert conversation here>

# Task:
- Go through the conversation history thoroughly
- For each rubric question think about your answer to each question carefully.
- Answer Y or N only to each rubric question.
- For Y answer, score your answer on a scale of 1-10 (low to high) to reflect how likely the expressed sentiment will impact the user's overall satisfaction/dissatisfaction with the interaction. For N answer, score 0.
- Only provide ONE most confident answer to each question.
- You *MUST* output your answers to all 10 questions provided in each rubric.
- Generate a table within <RESULTS></RESULTS> tags that summarizes your answers and scores to *ALL* 20 questions (10 questions in each rubric), using a Markdown table structure with rubric questions as the header (satisfaction rubric comes first, followed by dissatisfaction rubric).
- For each rubric question, output your answers and scores using the format "Y/N-SCORE".

# Answers

<RESULTS>

| The user thanks or compliments the AI agent for its responses.|The user continues to ask follow-up questions on the same or different topics. | The user does not express any confusion, frustration, or dissatisfaction with the AI agent's answers. | The user learns something new or useful by indicating curiosity and satisfaction with the information provided. | The user follows the AI agent's suggestions or instructions when applicable. | The user uses positive feedback words (e.g., excellent, amazing) or emojis, indicating enthusiasm and enjoyment of the conversation. | The user shares personal details or opinions with the AI agent when appropriate.|The user jokes with or challenges the AI agent in a friendly manner when suitable. | The user acknowledges or confirms that they understood or agreed with the AI agent's explanations when relevant. | The user ends the conversation on a positive note without asking for more information or assistance. | The user explicitly expresses dissatisfaction, frustration, annoyance, or anger with the AI agent's response or behavior. | The user repeatedly asks the same or similar questions or requests.|The user corrects the AI agent's mistakes or inaccuracies in its information or output. | The user changes the topic abruptly without acknowledging or following up on the previous one(s). | The user does not respond to the AI agent's questions, suggestions, feedback requests, etc. | The user has unrealistic expectations of what the AI agent can do and does not accept its limitations or alternatives. | The user implies that their query was ignored completely or that the response did not address their intent/goal at all. | The user perceives a decline in quality of service compared to previous experience with other agents/tools, etc. | The user wants more specific/useful information than what is provided by the AI agent. |The user feels that there is a mismatch between their preferred style and what is provided by the AI agent. |

|---|---|---|---|---|---|---|---|---|---|---|---|---|---|---|---|---|---|---|---|